\documentclass{article}
\usepackage{slashed}
\usepackage[dvipsnames]{xcolor}
\usepackage{amsmath}

\def\beq{\begin{eqnarray}}
\def\eeq{\end{eqnarray}}
\def\bea{\begin{eqnarray*}}
\def\eea{\end{eqnarray*}}

\def\centeron#1#2{{\setbox0=\hbox{#1}\setbox1=\hbox{#2}\ifdim
\wd1>\wd0\kern.5\wd1\kern-.5\wd0\fi
\copy0\kern-.5\wd0\kern-.5\wd1\copy1\ifdim\wd0>\wd1
\kern.5\wd0\kern-.5\wd1\fi}}
\def\ltap{\;\centeron{\raise.35ex\hbox{$<$}}{\lower.65ex\hbox{$\sim$}}\;}
\def\gtap{\;\centeron{\raise.35ex\hbox{$>$}}{\lower.65ex\hbox{$\sim$}}\;}

\def\singleandhalfspaced{\baselineskip=\normalbaselineskip\multiply
    \baselineskip by 150\divide\baselineskip by 100}

\newcommand{\newc}{\newcommand}
\newc{\qbar}{{\overline q}}
\newc{\Kahler}{K\"ahler }
\newc{\deltaGS}{\delta_{\rm GS}}
\begin{document}
\begin{titlepage}
\begin{flushright}
%{\large hep-th/yymmnnn \\
{\large SCIPP 13/14\\
}
\end{flushright}

\vskip 1.2cm

\begin{center}

{\LARGE\bf Anomaly Mediation in Local Effective Theories}

\vskip 1.4cm

{\large  Michael Dine and Patrick Draper}
\\
\vskip 0.4cm
{\it Santa Cruz Institute for Particle Physics and
\\ Department of Physics,
     Santa Cruz CA 95064  } \\
\vskip 4pt

\vskip 1.5cm

\begin{abstract}
The phenomenon known as ``anomaly mediation" can be understood in a variety of ways.  Rather than an anomaly, certain
gaugino bilinear terms are required by local supersymmetry and gauge invariance (the derivation of these terms
is in some cases related to anomalies in scale invariance or $R$ symmetries).  
We explain why the gaugino bilinear is required in supersymmetric gauge theories with varying number of colors and flavors.  By working in the Higgs phase, gauging a flavor group, or working below the scale of gaugino condensation, each of these theories has a local effective description in which we can identify the bilinear term, establishing its necessity in the microscopic theory.
For example, in theories that exhibit gaugino condensation, the potential in the very low energy theory is supersymmetric precisely due to the relation between the nonperturbative superpotential and the gaugino bilinear terms.  Similarly, the gravitino mass appears from its coupling to the gaugino bilinear.

\end{abstract}

\end{center}

\vskip 1.0 cm

\end{titlepage}
\setcounter{footnote}{0} \setcounter{page}{2}
\setcounter{section}{0} \setcounter{subsection}{0}
\setcounter{subsubsection}{0}

%%%%%%%%%%%%%%%%%%%%%%%%%%%%%%%%%%%%%%%%%%%
%%%%%%%%%%%%%%%%%%%%%%%%%%%%
\singleandhalfspaced

\section{Introduction:  Anomaly Meditation in its Various Guises}
\label{introduction}

It has been known for some time that gaugino masses can arise in theories
of low energy supersymmetry breaking, even when there is no coupling of the
Goldstino (longitudinal component of the gravitino) to the gauge multiplet~\cite{dinemacintire,anomalymediation1,
anomalymediation2}. More precisely,
such masses occur in cases where they cannot be understood as arising from local terms in a supersymmetric
effective action; there are contributions to these masses even in theories in which no field with a non-zero
$F$ component couples to the appropriate $W_\alpha^2$.  These terms are gaugino bilinears
multiplying the superpotential, and indeed arise in the absence of
supersymmetry breaking for theories in AdS space.  This phenomenon has been dubbed ``anomaly mediation."  As explained
in~\cite{dineseiberganomalymediation}, however, there is no anomaly associated with these couplings.  Rather, these terms
are contact terms {\it required} by supersymmetry.  They can be understood in a variety of ways.  In some cases they
arise with suitable regulators~\cite{dinemacintire}.  They can also be obtained by manipulating the conformal compensator
of certain supergravity formalisms~\cite{anomalymediation1,anomalymediation2}.  In a more broadly-applicable field theoretic approach, they can be associated
with supersymmetric completions of non-local
couplings~\cite{baggeretal,dineseiberganomalymediation,thaler,D'Eramo:2013mya}.  In certain circumstances, they are associated with local interactions~\cite{dineseiberganomalymediation}.  

To further elucidate this phenomenon, in the present paper we extend the arguments of~\cite{dineseiberganomalymediation} to
broader classes of theories.  In particular, in Ref.~\cite{dineseiberganomalymediation}, a simple argument for the gaugino mass was
given for $U(1)$ theories, with charged fields $\phi^\pm$, and superpotential independent of $\phi^+ \phi^-$,
\beq
W = W_0.
\eeq
In this case, the theory
has a flat direction  (neglecting gravitational strength interactions) with
\beq
\vert \phi^+ \vert = \vert \phi^- \vert = v.
\eeq
In this  direction, the gauge group is Higgsed, and there is one light chiral field, which we can take to be the gauge invariant combination
$\phi^+ \phi^-$.  We can study the 1PI action for the (heavy) gauge field.  This action is local, since there are no
couplings of the form $VLL$, $V$ denoting the vector fields, and $L$ the light chiral field.   In the
global limit, at one loop, the effective action includes a term:
\beq
\Gamma_{1PI} = -{1 \over 32 \pi^2} \int d^2 \theta d^4x (\tau - \log (\phi^+ \phi^-)) W_\alpha^2.
\eeq 
In a supergravity theory, this  modification of the gauge coupling function, $f$ leads to a term:
\beq
{\cal L}_{\lambda \lambda} = 
\lambda \lambda \left ({\partial f \over \partial \phi_+} {\partial K \over \partial \phi_+^*} W_0^* + (\phi^+ \rightarrow \phi^-) \right )
\eeq
$$~~~~= {1 \over 16 \pi^2} W_0^* \lambda \lambda\;,$$
precisely the anomaly-mediated expression for this case.   Everything in this analysis is completely local.  The term
is also independent of $v$, and so survives into the unbroken phase. Some regulators
(Pauli Villars) generate this term automatically; using a regulator
which does not generate this term (as, for example, is the case for dimensional regularization), then it is necessary to add
it as a finite counter term.
The treatment here, first of the global limit and then of gravitational corrections, is justified by the smallness of the gravitational coupling relative to $gv$\footnote{\label{fn1} In the 
unbroken phase, it has been shown in~\cite{Gripaios:2008rg} that infrared contributions in AdS space exactly cancel the contact term, leaving the gaugino massless if supersymmetry is unbroken. We will study the contact term but have in mind the physically relevant case where SUSY breaking lifts AdS to flat space, in which case the boundary contribution is not present.}.

While the analysis has the virtue of locality, and the field-independence of the contact term reproduces the contact term in the massless theory, it would be interesting to see the term arise in a Wilsonian context.  In this note, we consider more general
theories, surveying the
general case of non-Abelian theories with multiple flavors.  In an $SU(N)$ theory, for example,
if supersymmetry is unbroken or the breaking is small, the behavior of the theory and even
the questions one asks, are sensitive to the number of flavors.  For $N_f = N-1$, in the Higgs
phase the 1PI action is local, and one can repeat the argument of~\cite{dineseiberganomalymediation}.
For $N_f < N-1$, the theory exhibits gaugino condensation
and the ``anomaly mediated" interaction is responsible for a term in the potential {\it required} by local supersymmetry.
For $N_f \ge N$, the 1PI action is no longer local.  This can be addressed by gauging some of the flavor symmetry.
In general, there remain some light $U(1)$ gauge bosons, which can be described by a  Wilsonian effective action.
For these, local supersymmetry requires gaugino contact terms with coefficients precisely of the anomaly-mediated form.

We note that recently Refs.~\cite{thaler, D'Eramo:2013mya} have analyzed anomaly mediation in detail and demonstrated that there are several distinct mechanisms at work, physically separable by the goldstino coupling to the supercurrent.  The gaugino mass contributions we will identify correspond to pure gravitino mediation in the language of~\cite{thaler, D'Eramo:2013mya}, although as we will see the issue of the goldstino coupling is somewhat subtle. The masses we study are bulk AdS masses, which are equivalent to the flat-space masses when the SUSY-breaking is sequestered from the visible sector (see also footnote~\ref{fn1}.) We would also like to draw attention to Ref.~\cite{Sanford:2010hc}, which demonstrated the equivalence of the original discussion of anomaly mediation in~\cite{anomalymediation1} and the perspective of~\cite{dineseiberganomalymediation} and this work\footnote{For a differing interpretation that disagrees with ours, see~\cite{deAlwis:2008aq,deAlwis:2012gr}.}.

In the next section, we consider $U(1)$ theories with multiple chiral fields.  These fields already raise issues of locality,
which can be solved by introducing additional $U(1)$s.  Once the 1PI action is local, we directly recover the
anomaly-mediated expression.  In section \ref{non-Abelianlargenf}, we turn to non-Abelian gauge theories, in which the gauge group is
completely Higgsed on the moduli space.  We will see that it is not always true that the effective 
action is local, and will deal with this by gauging some of the flavor symmetry.  In the resulting cases,
there are frequently unbroken $U(1)$ symmetries, and for these, the gaugino counterterm
can be understood in a Wilsonian language.  In section \ref{puregauge}, we consider theories with
$N_f< N-1$, again in the presence of 
a small constant in the superpotential, $W_0$.   At low energies, there is an additional, dynamical
contribution to the superpotential.  The corresponding correction to the potential, linear
in $W_0$, is generated by the
gaugino contact term.  We demonstrate how this arises both
for the $N_f = 0$ and $N_f \ne 0$ cases. We also study the term linear in the superpotential and bilinear in the gravitino, which
we will refer to as the ``gravitino mass term."  In the final section,
we summarize and conclude.

\section{$U(1)$ Theories with More Than one Flavor}
\label{u1multipleflavors}

We can complicate the problem of~\cite{dineseiberganomalymediation} by considering {\it two pairs} of charged fields coupled to the vector multiplet.  We now have a more
interesting set of flat directions, but the main point is that the vector multiplet couples to a pair of massless
charged chiral fields as in the
Coulomb phase.  As a result, the 1PI action is no longer local, and we cannot infer the form of the full action so simply.  The
1PI action includes the local piece of~\cite{dineseiberganomalymediation}, as well as a non-local
term of the form described in~\cite{baggeretal}.  We can
modify the system so as to achieve locality in two ways.  First, we can introduce a mass for one of the fields.  This eliminates one of the flat directions, and leaves
us with the result for the contact term as in the single pair case.  This is just the statement that massive fields do not contribute to the
gaugino contact term.  But a more interesting infrared regulator is provided by introducing an additional $U(1)$, under which
we can take the fields to be:
\beq
\phi^+_+,\;\phi^-_+,\; \chi^+_-,\;\chi^-_-,
\eeq
where superscripts refer to the charge under the first $U(1)$, subscripts charge under the second.
Then, for example, there is a flat direction with all fields having equal vev.  There are two massless fields, which one can think of as:
\beq
L_1 = \phi^+_+ \chi^-_-\;,~~~~L_2 = \phi^-_+ \chi^+_-.
\eeq
The two heavy fields are the Higgs fields of the two $U(1)$s.  
There are no couplings of the form $V_a L_i L_j$, where
$V_a$ denote the two vector fields, and as a result,
the contributions to the 1PI action are infrared finite.  Indeed, the gauge coupling function has the form:
\beq
f={ g^{-2}(M)} - {1 \over 8 \pi^2} \log (L_1 L_2/M^4).
\eeq
From this we obtain the $\lambda \lambda$ contact term:
\beq
{\cal L}_{\lambda \lambda} = {1 \over 8 \pi^2}  W_0^*+{\rm c.c.}.
\eeq
This can be generalized to $N$ pairs of fields, again yielding the expected anomaly-mediated form. The $U(1)$ case is rather simple, but we will see similar phenomena in non-Abelian theories.

\section{$SU(N)$ with $N_f \ge N-1$}
\label{non-Abelianlargenf}

It is interesting to extend this analysis to non-Abelian theories.   In particular, consider an $SU(N)$ theory with $N_f$ flavors.  If $N_f \ge N -1$,
there are directions in which the gauge symmetry is completely Higgsed.  But, except for the case $N_f = N-1$, there are
massless fields with couplings to the vector fields, as in the $U(1)$ example above, and the 1PI effective action is non-local.  In the case $N_f = N-1$,
we can compute the effective action, which now contains:
\beq
\Gamma = -{1 \over 32 \pi^2} \int d^2 \theta \left (\tau - {2N+1 \over (2N-2) } \log (\det(\bar Q Q))\right ) W_\alpha^2.
\eeq 
This leads to a term:
\beq
{\cal L}_{\lambda \lambda} = {(2N+1) \over 16 \pi^2} \lambda \lambda W_0
\eeq
as expected from the usual anomaly-mediated formula.

For $N_f > N-1$, the 1PI action contains local terms and non-local interactions (in superspace).  Both contribute
to the gaugino bilinear.  But we can regulate the infrared as in the Abelian case by gauging some of the flavor symmetry.
In particular, we can gauge the $SU(N_F)$ symmetry.  Using the symmetries, the $Q$ and $\bar Q$ fields can
be brought to the form
with
\beq
\vert \bar v_a \vert^2 = \vert v_a \vert^2\;.
\eeq

Consider, first, the case $N_f = N$.  Then the gauge symmetry is broken, for general $v_a$, to
$U(1)^{N-1}$.  There are $N$ light chiral multiplets in these vacua; there are no couplings of the form $V L L$,
so the effective action {\it for the light vector multiplets} is local.  For, say, the $SU(N)$, there is a term:
\beq
\Gamma = -{1 \over 32 \pi^2} \int d^2 \theta \left (\tau - {2N \over (2N)} \log (\det(\bar Q Q)) \right ) W_\alpha^2.
\label{sunaction}
\eeq 
This leads to the anomaly-mediated mass term,
\beq
{\cal L}_{\lambda \lambda}= {1 \over 2}{2N \over 16 \pi^2} \lambda \lambda W_0^*\;.
\label{sunbilinear}
\eeq
There is a similar gaugino bilinear contact term for the
$SU(N_F=N)$ gauge group.  We find it remarkable that in this case anomaly mediation
is described by a Wilsonian effective action.

There are, however, two puzzles regarding Eq.~(\ref{sunbilinear}). The first is that the scales set by the $v_a$ are supersymmetric thresholds, and typically supersymmetric thresholds shift the gaugino masses so that they satisfy 
\begin{align}
m_\lambda\propto\beta^{IR}\;, 
\end{align}
where $\beta^{IR}$ is the beta function of the infrared effective theory. Therefore, below the scale of the smallest vev, we might expect $N-1$ gauginos with vanishing anomaly-mediated masses, because their beta functions vanish in the IR theory. But we have just found Eq.~(\ref{sunbilinear}) contributes to the light gaugino masses.

The resolution of the puzzle can be understood both in the effective and microscopic theories. From the IR perspective, the mechanism is that of deflected anomaly mediation~\cite{Pomarol:1999ie}. For simplicity, let us take $N=2$, so that there are two light singlets and one light $U(1)$. The singlets (call them $X_i$) have couplings to $W_\alpha^2$:
\begin{align}
(\bar{Q}Q)_{ii}&\rightarrow v_i^2+v_i (\delta Q_i +\delta\bar{Q}_i) + \dots\;\nonumber\\
&\equiv v_i^2+v_i X_i + \dots\;,\nonumber\\
\int d^2 \theta \log (\det &(\bar Q Q))  W_\alpha^2  \rightarrow \int d^2 \theta (X_i/v_i)W_\alpha^2+\dots
\label{XW2}
\end{align}
The $X_i$ also have linear terms in the~\Kahler potential:
\begin{align}
K=Q^\dagger_iQ_i+\bar{Q}^\dagger_i\bar{Q}_i\rightarrow v_i X_i + {\rm c.c.} + \dots
\end{align}
The linear terms generate an $F_{X_i}$,
\begin{align}
F_{X_i}\sim v_i W_0\;,
\end{align}
which provides a gaugino mass through the coupling of Eq.~(\ref{XW2}). Therefore, from the IR perspective, the spectrum simply does not look anomaly-mediated;  rather, it looks gauge-mediated, with masses mysteriously correlated with the cosmological constant.

In the microscopic theory, the question is why the supersymmetric threshold corrections did not keep the light gaugino masses on the anomaly-mediated trajectory. One way to understand this is to recall how decoupling works in the simple $U(1)$ theory with a large superpotential mass for the charged matter fields,
\begin{align}
W=W_0+m\phi_+\phi_-\;.
\end{align}
As reviewed in~\cite{dineseiberganomalymediation}, the $F$-term scalar potential contains $B$-term interactions between $W_0$ and $\phi_\pm$:
\begin{align}
V\sim - m m_{3/2} (\phi_+\phi_-+{\rm c.c.})\;.
\label{Bterm}
\end{align}
When $\phi_{\pm}$ are integrated out at $m$, $B$-term insertions in the threshold correction to the gaugino mass generate the $m$-independent shift
\begin{align}
m_\lambda = \frac{1}{16\pi^2}m_{3/2}\;.
\end{align}
This contribution exactly cancels an equal-and-opposite contribution from the regulator (e.g. Pauli-Villars fields). However, the essential point is that in supergravity, {\it $B$-terms are not generated by $D$-terms}. Supersymmetric mass thresholds that originate in the $D^2$ part of the supergravity potential, as in the microscopic $N=N_f$ theory on the Higgs branch, do not provide a threshold correction to $m_\lambda$ at leading order in $W_0$. Therefore, in such cases the light and heavy gauginos receive equal masses from the regulator, proportional to the beta function of the UV theory.

The second puzzle concerns the goldstino couplings. The $X W_\alpha^2$ interaction contains a goldstino coupling, which is not expected in strict gravitino mediation~\cite{thaler, D'Eramo:2013mya}. The resolution is that the microscopic theory breaks SUSY by a small amount $F_Q\sim v W_0.$ This $F$-term does not give masses to the gauginos at first order in $W_0$ and zeroth order in $v/M_p$ (therefore gravitino mediation really is the source of the masses), but it does imply that there will be a goldstino effective coupling with coefficient $m_\lambda / F_Q \sim 1/v$. In the IR effective theory, the gaugino counterterm and the goldstino coupling combine to give the local effective superfield coupling $X W_a^2$. As explained above, in the IR the gaugino mass is provided by $F_X/v \sim W_0$, even though diagrams involving $F_Q$ were not the source of the mass in the UV theory.

As an aside, it should be noted that there is a gauge-invariant description of the unbroken symmetry, in terms of operators
\beq
W_{a \bar b}  Q^a_f Q^{\bar b}_{\bar f} \delta^{f \bar f}
\eeq
and additional operators constructed with $W_{f,\bar f}$ and $\epsilon$ tensors.  

Now consider $N_f = N+1$.  Here the low energy group is $U(1)^{N-1}$.  There are only $N$ light chiral fields, for general
choice of the $Q$ and $\bar Q$ vev's.  Repeating the analysis leading to Eq.~(\ref{sunbilinear}) leads to the
bilinear contact term expected from the anomaly mediation formula.  

In the case $N_f > N+1$, the low energy gauge group is $U(1)^{N-1} \times SU(N_f-N)$.  There are no
chiral fields transforming under the $SU(N-N_F)$.  In this sector, gaugino condensation occurs and there is a mass
gap; again the very low energy theory consists of a set of massless $U(1)$s and $N$ chiral fields.  Constructing the effective action yields, for each of the $U(1)$s,
\beq
\Gamma = -{1 \over 32 \pi^2} \int d^2 \theta (\tau -{(3N-N_f )\over 2N_f} \log (\det(\bar Q Q))) W_\alpha^2.
\label{sunaction1}
\eeq 
where the determinant is in the $SU(N)$ indices.   Again, we obtain the anomaly-mediated formula, proportional to the beta function of the UV theory above the scale of the $v_a$. The splitting between the gaugino and gauge boson masses is the same for the light and heavy gauginos for the reason discussed previously: from the point of view of the IR effective theory, there is a singlet coupling to $W_\alpha^2$ that takes the light gauginos off the anomaly-mediated track; from the point of view of the microscopic theory, the gaugino masses are all anomaly-mediated, but thresholds that change the beta functions do not change the masses.

One can ask about the gaugino condensate.  In particular, this should lead to an effective superpotential at low
energies, and correspondingly there should be a negative contribution to the cosmological constant, and a
mass-like term for the gravitino.  Such terms, in this context and for theories with $SU(N)$ gauge groups
and $N_f < N-1$, will be the subject
of the next sections.

\section{Low Energy Effective Lagrangian: Theories of Gluino Condensation}
\label{puregauge}

In theories with $0 \le N_f < N-1$, the gauge symmetry is not completely broken at general
points on the classical moduli space; in the remaining theory, there is a mass gap and gaugino
condensation occurs.  At lower energies, one should have a supergravity theory, with chiral fields if $N_f > 0$.
We first demonstrate that, in a pure gauge theory (i.e. no chiral fields), accounting for the
full supergravity potential {\it requires} the presence of precisely the contact terms expected
from anomaly mediation.  We then consider the more general case.

\subsection{Pure Gauge Theory:  Gaugino Condensation}
In a pure gauge theory coupled to supergravity, with no other degrees of freedom, it makes sense to consider an effective low energy theory well below the scale of the confining gauge theory.
If we include in the pure gauge theory a constant superpotential $W_0$, in the low energy theory, we should be able to identify the term
\beq
-3 W_{np} W_0^* + {\rm c.c.}
\label{WnpW0}
\eeq
in the potential, where $W_{np}$ is the nonperturbatively generated superpotential, as well as the term
\beq
(W_0 + W_{np}) \psi_\mu \sigma^{\mu \nu} \psi_\nu
\eeq
for the gravitino.

The first of these terms requires adding the ``anomaly-mediated" gaugino mass counterterm to the high-energy theory:
\beq
{\cal L}_{\lambda \lambda}={1\over 2}{b_0 \over 16 \pi^2} W_0^* \lambda \lambda\;,\;\;\;\;\;\;b_0\equiv 3N
\eeq
From the relation between the nonperturbative superpotential in the effective theory and the gaugino condensate,
\beq
\langle \lambda \lambda \rangle = -{32 \pi^2 \over N} W_{np},
\eeq
we see that we immediately recover Eq.~(\ref{WnpW0}). Note that the factor of $3$ in $b_0$ is crucial.

There should be one other term linear in the dynamical superpotential.  The term
\beq
{\cal L}_{\psi\psi}\sim W_{np} \psi_\mu \sigma^{\mu \nu} \psi_\nu
\eeq
will be discussed in section \ref{gravitinomass}.

Finally, it should be stressed that the action under consideration is {\it Wilsonian}; we have properly integrated out massive, high energy degrees of freedom, to obtain an effective action for the light fields (the graviton supermultiplet, as well as the Goldstino, in the case that supersymmetry is broken).  It should also be stressed that this analysis, in the spirit of~\cite{dineseiberganomalymediation}, can be viewed as a {\it derivation} of the anomaly mediated result for this theory.
If one works with a regulator which does not generate the gaugino bilinear, supersymmetry {\it requires} that one add
it as a counterterm.

\subsection{$N_f < N-1$}
\label{non-Abeliansmallnf}

For $0 < N_f < N-1$, on the moduli space the gauge group is not completely higgsed.  In general, the low energy
theory is a pure gauge theory with $N-N_f$ colors, along with a set of light chiral multiplets (pseudomoduli), neutral
under the gauge group.  The non-perturbative superpotential takes the form: 
\beq
W_{np} = {\Lambda^{3N-N_f \over N-N_f} \over \det (\bar Q Q)^{1 \over N-N_f}}.
\label{wnp}
\eeq
As a result, the {\it potential}, at low energies, 
\beq
V = \vert {\partial W \over \partial Q_f} +{ \partial K \over \partial Q_f} W\vert^2  + (Q \rightarrow \bar Q) - 3 \vert W \vert^2
\eeq
contains a term:
\beq
W_0^* W_{np} (-3 -{2 N_f \over N-N_f})= {-3N + N_f  \over N-N_f} W_0^* W_{np}.
\eeq
Gaugino condensation in the $SU(N-N_f)$ group is the origin of $W_{np}$,
\beq
W_{np} = -{N-N_f \over 32 \pi^2} \langle \lambda \lambda \rangle\;,
\eeq
so the microscopic theory must contain
\beq 
{\cal L}_{\lambda \lambda} = {3N - N_f \over 32 \pi^2} \lambda \lambda W_0^* + {\rm c.c.}.
\eeq
This is the anomaly-mediated gaugino mass counterterm. It is proportional to the beta function of the microscopic theory, instead of the beta function of the low-energy effective theory as one might otherwise expect, for the reasons discussed in Section 3.
Again, it is worth stressing that this analysis is strictly Wilsonian, and it is again clear that supersymmetry and gauge invariance
{\it require} the presence of the gaugino bilinear.

\subsection{Small Quark Masses}

A further test of these ideas is provided by considering theories with small quark masses.  
For  $N_f <N$ and small quark masses, instead of describing a system with a pseudomoduli space, one has supersymmetric
(AdS) vacuua.
Taking for the (classical) superpotential:
\beq
W = m_f \bar Q_f Q_f
\eeq
with $m_f$ small compared to the dynamical scale of the theory, one can compute a dynamical contribution to the superpotential,
as in Eq.~(\ref{wnp}).  At the supersymmetric stationary point,
\beq
m_f \bar Q_f Q_f = {1 \over N-N_f } W_{np} \;\;\; \rm (no~sum~over~f)
\eeq
Then
\beq
\langle W \rangle = ({N_f  \over N-N_f} + 1) W_{np} = -{N \over 32 \pi^2} \langle \lambda \lambda \rangle.
\eeq
Now turning on a small constant in the superpotential, the term linear in $W_0$ in the microscopic theory  (before including the counterterm) is:
\begin{align}
{\cal L}_{W_0} &=  W_0^* \left(-3 m_f \bar Q_f Q_f  + {\partial W \over \partial Q_f}{\partial K \over \partial Q_f^*} +  {\partial W \over \partial \bar{Q}_f}{\partial K \over \partial \bar{Q}_f^*}\right) + {\rm c.c.}\nonumber\\
&=W_0^*\left(-m_f\bar{Q}_f Q\right)+{\rm c.c.}
\end{align}
Adding the counterterm,
\beq
{\delta \cal{L}}_{\lambda \lambda} = {3N - N_f \over 32 \pi^2}W_0^* \lambda \lambda + {\rm c.c.}
\eeq
yields precisely
\beq
\delta V = -3 W_0^* W + {\rm c.c.}.
\eeq
So again we see that supersymmetry requires the counterterm.

\section{The gravitino bilinear}
\label{gravitinomass}

In the theories with $N_f < N-1$, there should be an additional term in the low energy effective action linear in $W_{np}$:
\beq
{\cal L}_{\psi\psi}=-e^{K/2} W \psi_\mu \sigma^{\mu \nu} 
\psi_\nu.
\label{gravitinomassterm}
\eeq
This term requires a strong coupling analysis and we will content ourselves with some heuristic remarks.
First, we choose a gauge
\beq
\bar \sigma^\mu \psi_\mu = 0.
\eeq
(We are using the spinor conventions of Wess and Bagger).  Then Eq.~(\ref{gravitinomassterm}) becomes
\beq
{\cal L}_{\psi\psi}=-e^{K/2}{1 \over 2} W \psi^\mu \psi_\mu.
\eeq
Roughly
speaking, to account for this coupling, we are looking for terms in the action of the form
\beq
{\cal L}\sim \bar{\lambda} \bar{\lambda} \psi^\mu \psi_\mu
\eeq
These can arise from the two sources:
\beq
{\cal L}_{\psi\psi\lambda\lambda}={1 \over 8} {\rm Re} f  \psi_\mu \sigma^{\rho \sigma} \sigma^\mu \bar{\lambda} (\psi_\rho \sigma_\sigma \bar{\lambda} - \psi_\sigma \sigma_\rho \bar{\lambda})+{\rm h.c.}
\label{psipsill}
 \eeq
and
\beq
{\cal L}_{\psi \lambda F} =
{i \over 4} {\rm Re} f  \psi_\mu \sigma^{\rho \sigma} \sigma^\mu \bar{\lambda}   F_{\rho \sigma}+{\rm h.c.}\;,
\eeq
where $f$ is the gauge coupling function.
Eq.~(\ref{psipsill}) reduces to
\beq
{\cal L}_{\psi\psi\lambda\lambda}=-{1 \over 4}{\rm Re} f  \langle\bar{\lambda}\bar{\lambda}\rangle\psi\psi
\label{term1}
\eeq
If we work perturbatively to second order in ${\cal L}_{\psi \lambda F}$,
the exchange of the gauge field also yields an effective $\bar{\lambda} \bar{\lambda}\psi \psi$ interaction.
At large $q^2$, $q^2 \gg \Lambda^2$ for the gravitino, we can give an (again heuristic) justification of this
calculation.  In the $\psi\psi$ two-point function, there is a term
\beq
\langle  {\rm Re} f  \psi_\mu(x) \sigma^{\rho \sigma} \sigma^\mu \lambda(x)   F_{\rho \sigma}(x){\rm Re} f  \psi_\nu(y) \sigma^{\gamma \delta} \sigma^\nu \lambda(y)   F_{\gamma \delta}(y) \rangle \;.
\eeq
Taking $x \rightarrow y$ and using the operator product expansion, one obtains a contribution
\beq
{1 \over 8} {\rm Re}f  \langle \bar{\lambda} \bar{\lambda} \rangle \psi \psi\;.
\label{term2}
\eeq
Combining Eqs.~(\ref{term1}) and~(\ref{term2}), 
\beq
m_\psi={1\over 4}{\rm Re}f\langle\bar{\lambda}\bar{\lambda}\rangle\;.
\label{mpsi}
\eeq
We need to understand at what scale to evaluate the function $f$, and to determine its value.  Again, we can
give a heuristic computation, in this case using the Veneziano-Yankielowicz  Lagrangian.
We present the Lagrangian in a somewhat different form than is standard, which allows more immediate
connections to the gaugino contact terms.
We take
$S= W_\alpha^2$
and treat it as an elementary chiral field.  For $W(S)$, we take:
\beq
W = {1 \over 4} (\tau S + \frac{N}{8\pi^2} S \log S/M^3)
\label{unconventionalvysuperpotential}
\eeq
where $M$ is a UV cutoff and the microscopic theory is pure $SU(N)$.  This is slightly different than the original presentation by VY, but, like theirs, it properly respects the
discrete symmetries (and an $R$ symmetry under which the spurion, $\tau$, transforms).  In this form, the action
has precisely the structure of the holomorphic effective action,
evaluated at the scale $S$.
The stationary point is:
\beq
S = \Lambda^3 e^{-1}\;,\;\;\; \Lambda = M e^{-8 \pi^2 \tau/N}\;.
\eeq
At the minimum,
\beq
\langle W \rangle = -{N \over 32 \pi^2} \langle S \rangle.
\eeq
At this point, $f = -{N \over 8 \pi^2}.$
So Eq.~(\ref{mpsi}) indeed becomes
\beq
m_\psi=\langle W\rangle\;.
\eeq

The VY superpotential is often presented in a different fashion, and understanding the connections is instructive.   In particular,
one often sees
\beq
W = N (S \log(S/\Lambda^3) + S).
\eeq
This leads to
\beq
\langle W \rangle = -N \langle S \rangle
\eeq
in which case $S$ must be identified with ${1 \over 32 \pi^2} W_\alpha^2$.  Instead, we consider
\beq
W = {N \over 8 \pi^2} (S \log(S/\Lambda^3) + mS)
\eeq
With $\Lambda = M e^{-{8 \pi^2 \tau \over N}} $
this is the superpotential of Eq.~(\ref{unconventionalvysuperpotential}), with a change of the coefficient of the $S$ term.  This does not alter
$\langle W\rangle$, which is independent of $\tau$, and can therefore be viewed as a finite coupling redefinition (change of scheme).  The basic results of this section are, from this perspective,
scheme independent.

\section{Conclusions}

Surveying $SU(N)$ gauge theories with different numbers of flavors provides insight into the necessary presence of bilinear gaugino counterterms in the Lagrangian
of supersymmetric
theories, the phenomenon known as ``anomaly mediation."  We have extended the arguments of~\cite{dineseiberganomalymediation} that such counterterms, rather
than indicating an anomaly, are {\it required} by local supersymmetry. By examining phases of the theories where local, low energy effective actions (often Wilsonian) are available, we have shown that the gaugino bilinears or their remnants are easily identified.

\vskip 1 cm
\noindent
{\bf Acknowledgements:}
We thank Francisco D'Eramo and Jesse Thaler for useful discussions. This work supported in part by the U.S. Department of Energy.

\newpage
\bibliographystyle{unsrt}
\bibliography{dinerefs}{}

\end{document}